\documentclass[aps,prb,twocolumn,superscriptaddress]{revtex4-1}

\newcommand{\figurescale}{1}
\usepackage{verbatim}
\usepackage{amsmath}
\usepackage[utf8]{inputenc}
\usepackage[pdftex]{graphicx}
\usepackage[separate-uncertainty=true]{siunitx}
\DeclareSIUnit{\rpm}{rpm}
\usepackage[colorlinks=true,urlcolor=blue,linkcolor=blue,citecolor=blue]{hyperref}
\usepackage[usenames,dvipsnames]{xcolor}
\usepackage[T1]{fontenc}
\usepackage{placeins}
\usepackage{nicefrac}
\usepackage{braket}
\usepackage{pdfcomment}
\usepackage{xcolor}
\usepackage{braket}
\usepackage[normalem]{ulem}

\usepackage{lineno}

\begin{document}

\title{Atomistic defect states as quantum emitters in monolayer MoS$_2$}
%
\author{J.~Klein}\email{julian.klein@wsi.tum.de}
\affiliation{Walter Schottky Institut and Physik Department, Technische Universit\"at M\"unchen, Am Coulombwall 4, 85748 Garching, Germany}
\affiliation{Nanosystems Initiative Munich (NIM), Schellingstr. 4, 80799 Munich, Germany}
\author{M.~Lorke}
\affiliation{Institut für Theoretische Physik, Universität Bremen, P.O. Box 330 440, 28334 Bremen, Germany}
\affiliation{Bremen Center for Computational Materials Science, University of Bremen, Am Fallturm 1, 28359 Bremen, Germany} 
\author{M.~Florian}
\affiliation{Institut für Theoretische Physik, Universität Bremen, P.O. Box 330 440, 28334 Bremen, Germany}
\author{F.~Sigger}
\affiliation{Walter Schottky Institut and Physik Department, Technische Universit\"at M\"unchen, Am Coulombwall 4, 85748 Garching, Germany}
\affiliation{Nanosystems Initiative Munich (NIM), Schellingstr. 4, 80799 Munich, Germany}
\author{J.~Wierzbowski}
\affiliation{Walter Schottky Institut and Physik Department, Technische Universit\"at M\"unchen, Am Coulombwall 4, 85748 Garching, Germany}
\author{J.~Cerne}
\affiliation{Department of Physics, University at Buffalo, The State University of New York, Buffalo, New York 14260, USA}
\author{K.~M\"uller}
\affiliation{Walter Schottky Institut and Physik Department, Technische Universit\"at M\"unchen, Am Coulombwall 4, 85748 Garching, Germany}
\author{T.~Taniguchi}
\affiliation{National Institute for Materials Science, Tsukuba, Ibaraki 305-0044, Japan}
\author{K.~Watanabe}
\affiliation{National Institute for Materials Science, Tsukuba, Ibaraki 305-0044, Japan}
\author{U.~Wurstbauer}
\affiliation{Walter Schottky Institut and Physik Department, Technische Universit\"at M\"unchen, Am Coulombwall 4, 85748 Garching, Germany}
\affiliation{Nanosystems Initiative Munich (NIM), Schellingstr. 4, 80799 Munich, Germany}
\author{M.~Kaniber}
\affiliation{Walter Schottky Institut and Physik Department, Technische Universit\"at M\"unchen, Am Coulombwall 4, 85748 Garching, Germany}
\affiliation{Nanosystems Initiative Munich (NIM), Schellingstr. 4, 80799 Munich, Germany}
\author{M.~Knap}
\affiliation{Department of Physics and Institute for Advanced Study, Technical University of Munich, 85748 Garching, Germany}
\author{R.~Schmidt}
\affiliation{Max-Planck-Institut f\"ur Quantenoptik, 85748 Garching, Germany}
\author{J.~J.~Finley}\email{finley@wsi.tum.de}
\affiliation{Walter Schottky Institut and Physik Department, Technische Universit\"at M\"unchen, Am Coulombwall 4, 85748 Garching, Germany}
\affiliation{Nanosystems Initiative Munich (NIM), Schellingstr. 4, 80799 Munich, Germany}
\author{A.~W.~Holleitner}\email{holleitner@wsi.tum.de}
\affiliation{Walter Schottky Institut and Physik Department, Technische Universit\"at M\"unchen, Am Coulombwall 4, 85748 Garching, Germany}
\affiliation{Nanosystems Initiative Munich (NIM), Schellingstr. 4, 80799 Munich, Germany}
%

%
\date{\today}
%
%
\begin{abstract}
\textbf{Quantum light sources in solid-state systems are of major interest as a basic ingredient for integrated quantum device technologies. The ability to tailor quantum emission through deterministic defect engineering is of growing importance for realizing scalable quantum architectures. However, a major difficulty is that defects need to be positioned site-selectively within the solid. Here, we overcome this challenge by controllably irradiating single-layer MoS$_{2}$ using a sub-nm focused helium ion beam to deterministically create defects. Subsequent encapsulation of the ion bombarded MoS$_{2}$ flake with high-quality hBN reveals spectrally narrow emission lines that produce photons at optical wavelengths in an energy window of one to two hundred $\SI{}{\milli\electronvolt}$ below the neutral 2D exciton of MoS$_{2}$. Based on ab-initio calculations we interpret these emission lines as stemming from the recombination of highly localized electron-hole complexes at defect states generated by the helium ion bombardment. Our approach to deterministically write optically active defect states in a single transition metal dichalcogenide layer provides a platform for realizing exotic many-body systems, including coupled single-photon sources and exotic Hubbard systems.
}
\end{abstract}
%
%
\maketitle
%
%
\section{Introduction}

Point defects are important for a variety of physical phenonema in semiconductors. For example, they provide a means to engineer the equilibrium free-carrier density, they can serve as quantum emitters, and may realize quantum bits for quantum information processors.\cite{Aharonovich.2016} A major challenge in several of these applications is that the defects need to be precisely positioned, which is particularly challenging for conventional three-dimensional semiconductors where defects are often buried deep in the bulk structure. A step to remedy this challenge is to reduce the physical dimension. Over the last decade it became possible to fabricate and manipulate atomically thin two-dimensional (2D) materials that offer intriguing electronic and optoelectronic properties.\cite{ajayan2016,Roldan.2017} In particular, the hexagonal transition metal dichalcogenides (TMDCs) such as MoS$_{2}$, MoSe$_{2}$, WS$_{2}$, and WSe$_{2}$ are excellent candidates for photonic applications due to their exceptionally strong light-matter interaction resulting from weak non-local dielectric screening.\cite{Mak.2010, Splendiani.2010, Florian.2017} The strong Coulomb coupling manifests itself in an exciton dominated spectral response and very large exciton binding energies.\cite{Mak.2010, Splendiani.2010, Ugeda.2014, Chernikov.2014, He.2014}

Beyond the response of mobile excitons, also quantum dot-like emission from localized excitons was recently demonstrated in WSe$_{2}$,\cite{Tonndorf.2015,Srivastava.2015,He.2015,Koperski.2015,Chakraborty.2015} GaSe,\cite{Tonndorf.2017} MoSe$_{2}$\cite{Branny.2016} and WS$_{2}$\cite{PalaciosBerraquero.2017} and the nature of the potential which localizes the excitons is not yet fully understood. The origin of such quantum emitters is hitherto considered to be caused by uncontrolled strain potentials that locally reduce the bandgap, thus, funneling the recombination of excitons via a discrete recombination center.\cite{Kern.2016,Branny.2017,PalaciosBerraquero.2017,Blauth.2018} However, strain potentials are challenging to control, which limits the applicability of such quantum emitters for a prospective integration into photonic circuits. \cite{Laucht.2012,Reithmaier.2015,Goodfellow.2014,Goodfellow.2015,Blauth.2017,Blauth.2018} In particular, scalable quantum technologies require the development of controlled approaches for the direct site-selective planar integration of quantum emitters with atomic scale resolution.

\begin{figure*}[!ht]
\scalebox{\figurescale}{\includegraphics[width=1\linewidth]{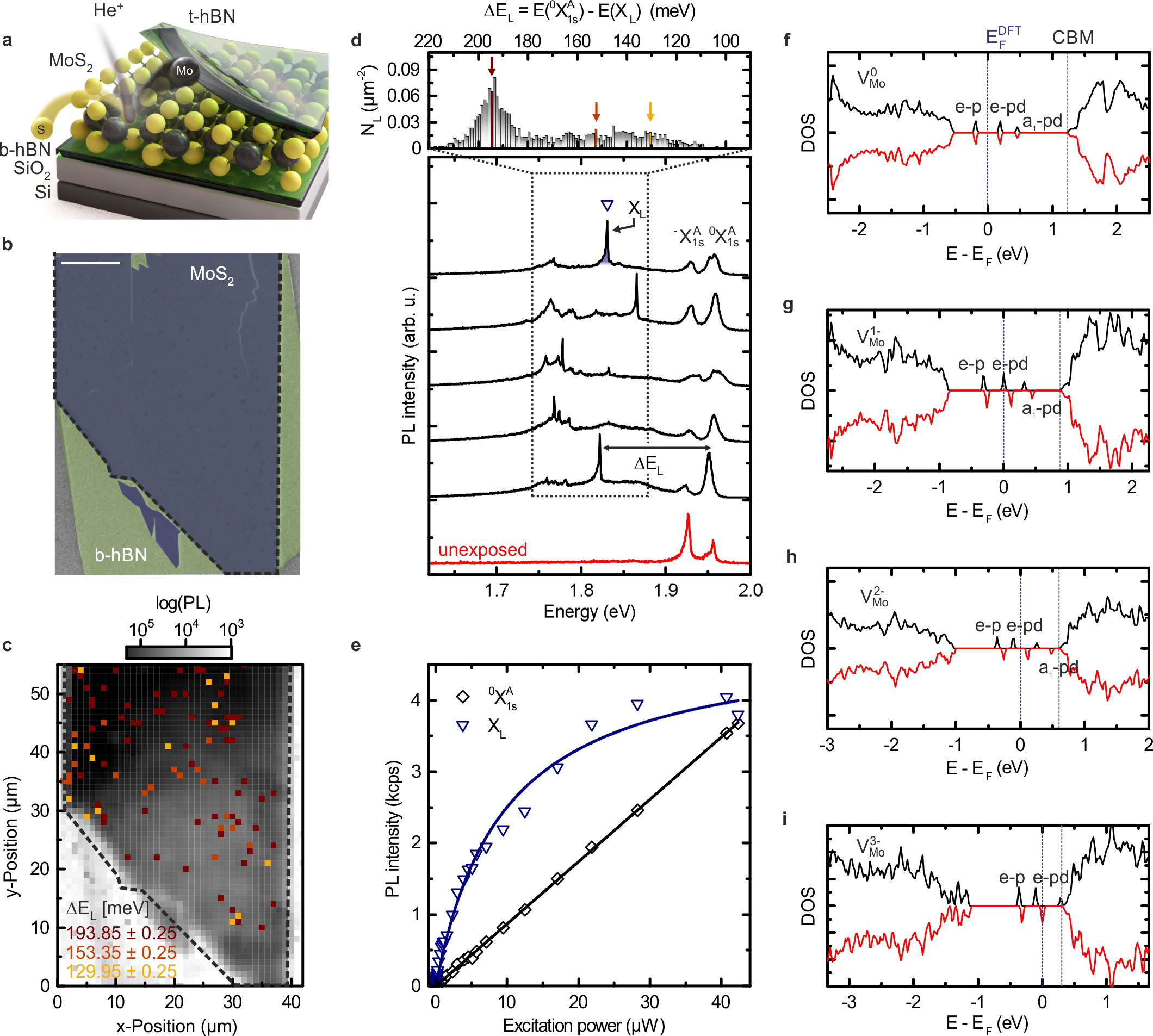}}
\renewcommand{\figurename}{Fig.}
\caption{\label{fig1}
\textbf{Deterministically induced single defect emitters in atomically thin MoS$_{2}$ realized by focused helium ions.} 
\textbf{a}, Schematic illustration of the bombarded MoS$_{2}$/hBN van der Waals heterostructure. 
\textbf{b}, Helium ion microscopy image taken at a dose of $\sigma = 2.2 \cdot 10^{12}\SI{}{\per\centi\meter\squared}$. The scale bar is $\SI{10}{\micro\meter}$. 
\textbf{c}, Spatially resolved and spectrally integrated photoluminescence mapping. The colored pixels depict the occurrence of emitters within $\SI{500}{\micro\electronvolt}$ energy bins (highlighted by colored arrows in \textbf{d}). 
\textbf{d}, Bottom panel: typical low-temperature ($\SI{10}{\kelvin}$) $\mu$-PL spectra of the bombarded (black) and non-bombarded (red) hBN/MoS$_{2}$/hBN heterostructure. The spectra of bombarded MoS$_{2}$ feature emission from mobile excitons $^{0}X^{A}_{1s}$ and trions $^{-}X^{A}_{1s}$, as well as single defect emission $X_{L}$ (open triangle) at lower energies. Top panel: a histogram of the emission energies detuned by $\Delta E_{L} = E(^{0}X^{A}_{1s}) - E(X_{L})$ below $^{0}X^{A}_{1s}$. 
\textbf{e}, Power dependence: $^{0}X^{A}_{1s}$ shows an expected linear power dependence, while the $X_{L}$ emission saturates for higher excitation powers (data for the emitter highlighted by open triangle in \textbf{d}). 
\textbf{f-i}, DFT calculated density of states (DOS) of the neutral V$_{Mo}^{0}$, single negatively V$_{Mo}^{1-}$, double negatively V$_{Mo}^{2-}$, and triple negatively charged V$_{Mo}^{3-}$ molybdenum vacancy. The DOS of V$_{Mo}^{0}$ shows doublet $e$-$p$, $e$-$pd$ and singlet $a_{1}$-$pd$ defect states inside the band gap. The $a_{1}$-$p$ singlet state is situated within the valence band. The more electrons are added to the vacancy the closer the defect states and the  DFT computed Fermi level energy $E^{DFT}_{F}$ shift to the conduction band minimum (CBM) because of the on-site Coulomb repulsion.
}
\end{figure*}

In this work, we propose and investigate an alternative route to engineer single defect emitters that are of atomic scale. Specifically, we demonstrate theoretically and experimentally that localized defects in two-dimensional layers create trapping potentials in which bound complexes are formed that emit in a range of about hundred meV below the free exciton line in MoS$_2$. Our experimental approach to systematically create such point defects is by helium ion bombardment of atomically thin MoS$_{2}$. Compared to strain engineering, a controlled bombardment with helium ions is superior due to the sub-nm beam size that is perfectly suited to structure 2D materials at the nanometer scale. \cite{Fox.2015} First attempts to generate optically active defects in TMDCs have been made by bombarding the host crystal with $\alpha$-particles \cite{Tongay.2013} and helium ions.\cite{Klein.2017} However, in these works a strong dynamical inhomogeneous linewidth broadening of $\sim \SI{40}{\milli\electronvolt}$ was imposed on the emission due to the fluctuating environment. Here, we demonstrate that the encapsulation of helium irradiated MoS$_{2}$ monolayers with hBN allows one to define single defect emitters with very narrow linewidths. The combination of ion-bombardment and subsequent encapsulation results in a hitherto unobserved new class of spatially localized and spectrally narrow defect emission in MoS$_{2}$ at very distinct energies on the order of $\sim \SI{100}{\milli\electronvolt} - \SI{220}{\milli\electronvolt}$ below the neutral exciton $^{0}X_{1s}^{A}$. Our spectroscopic observations are consistent with quantum emission that stems from defect centers in MoS$_{2}$, which serve as highly-localized exciton emission centers.
The interpretation is corroborrated by analyzing in detail the lineshape of the new emission peaks. This allows us to determine the interaction of the localized exciton with phonons which indicates that the emitters are localized on a nanometer scale.
Our study paves the way for inducing highly controllable single defect emitters in two-dimensional materials to create arrays of coupled single-photon sources,\cite{aharonovich_solid-state_2016} to study Anderson's orthogonality catastrophe,\cite{anderson_infrared_1967, schmidt_Universal_2018} and to realize lattices of excitons similarly to optical lattices for ultracold atomic gases.\cite{bloch_many-body_2008}

\section{Results}

The van der Waals heterostrutures studied in this work are iteratively stacked onto Si substrates covered with \SI{290}{\nano\metre} thermally grown SiO$_{2}$, followed by multilayered hBN, and a single layer of MoS$_{2}$. Typical hBN thicknesses vary between $\sim 10 - \SI{30}{\nano\meter}$ as confirmed by atomic force microscopy. The samples are transferred into a helium ion microscope, and are locally exposed to helium ions at a constant dose of $\sigma = 2.2 \cdot 10^{12} \SI{}{\per\centi\meter\squared}$, see Figure~\ref{fig1}a. Figure~\ref{fig1}b shows a helium-ion microscope (HIM) image of the bombarded van der Waals heterostructure. After the ion bombardment, the entire MoS$_{2}$ crystal is fully encapsulated by another multilayer hBN capping.

We perform low-temperature $\mu$-photoluminescence ($\mu$-PL) spectroscopy on the samples kept in a helium flow cryostat at a lattice temperature of 10 K. For excitation, we use a cw laser with a photon energy of $\SI{2.33}{\electronvolt}$ and a low excitation power (excitation power density) of less than $\SI{10}{\micro\watt}$ ($\SI{0.884}{\kilo\watt\per\centi\meter\squared}$). Figure~\ref{fig1}c shows the spatially resolved and spectrally integrated $\mu$-PL response as a false color representation. Representative luminescence spectra from five randomly selected positions of the exposed sample that show quantum emission are presented in the bottom of Fig.~\ref{fig1}d and compared to a spectrum of a pristine (unexposed) sample. All spectra reveal spectrally narrow neutral exciton $^{0}X^{A}_{1s}$ ($FWHM=\SI{8.3 \pm 2.5}{\milli\electronvolt}$) and charged exciton $^{-}X^{A}_{1s}$ ($FWHM=\SI{9.4 \pm 2.9}{\milli\electronvolt}$) emission whose inhomogeneous linewidth is reduced due to the encapsulation with hBN. \cite{Wierzbowski.2017,Ajayi.2017,Cadiz.2017,Florian.2017} Remarkably, in addition to the delocalized excitons, we observe spectrally sharper emission peaks $X_{L}$ ($FWHM\sim \SI{0.5}{\milli\electronvolt}-\SI{6}{\milli\electronvolt}$) in a  window of $\Delta E_{L} \sim \SI{100}{\milli\electronvolt} - \SI{220}{\milli\electronvolt}$ red-shifted from $^{0}X^{A}_{1s}$ (spectral range highlighted by a dotted box in Fig.~\ref{fig1}d) for helium bombarded samples. These features are superimposed on the so-called L-peak that is observed in single-layered MoS$_{2}$ \cite{Mak.2010,Splendiani.2010,Korn.2011} and has recently been attributed to the presence of sulphur vacancies.\cite{Carozo.2017} The sharp spectral features are spatially localized and require helium ion bombardment of MoS$_{2}$ and the subsequent encapsulation with hBN. Spectrally sharp peaks do not occur in non-bombarded hBN/MoS$_{2}$/hBN heterostacks (cf. bottom spectrum in Fig.~\ref{fig1}d), neither in bombarded MoS$_{2}$ without hBN encapsulation nor in bombarded hBN (cf. Supplementary Information).

To obtain further information on the origin of the sharp spectral emission, we statistically evaluate the energetic detuning $\Delta E_{L} = E(^{0}X^{A}_{1s}) - E(X_{L})$ of each localized emission peak $E(X_{L})$ with respect to $E(^{0}X^{A}_{1s})$. Figure~\ref{fig1}d, top, shows a histogram of the localized $X_{L}$ emission energies as a function of detuning $\Delta E_{L}$ for $\sim 3500$ emitters for one representative sample (cf. Supplementary Information). The histogram shows that the spectral emitters possess a continuum of energies in the range of $\Delta E_{L} \sim \SI{100}{\milli\electronvolt} - \SI{220}{\milli\electronvolt}$ below the 1s exciton. We also show the spatial occurence of emitters in three energy bins (cf. arrows in Fig.~\ref{fig1}d), overlayed in Fig.~\ref{fig1}c. So far, single-photon emitters have been found at sample edges or folds, where the bandstructure is strongly influenced by either the crystal's boundary or strain.~\cite{aharonovich_solid-state_2016} By contrast, we find spectrally sharp emitters homogeneously distributed all over the 2D sample; this is a direct consequence of the helium ion bombardment.

The sharp emission lines reveal a saturating behavior with a threshold power of $P_{sat} \sim \SI{10}{\micro\watt}$ (cf. Fig.~\ref{fig1}e), which is consistent with the assumption of saturating a finite density of a single defect state as fitted by

\begin{equation}
I \propto \frac{P}{P + P_{sat}}~.
    \label{eq:satpow}
\end{equation}

By contrast, the 2D exciton $^{0}X^{A}_{1s}$ shows the expected linear dependence with the excitation power.

Figures~\ref{fig1}f-i show the ab-initio calculated density of states (DOS) for a neutral $V^{0}_{Mo}$, as well as a single, double and triple negatively charged molybdenum vacancy ($V^{1-}_{Mo}$, $V^{2-}_{Mo}$ and $V^{3-}_{Mo}$). The vacancy states have a trigonal and mirror symmetry between the two layers of the S and the Mo layer.\cite{Noh.2014} In turn, two singlet ($a_{1}$) and two doublet ($e$) states are expected for all cases. The lowest $a_{1}$-$p$ [$e$-$p$] level stems from the six adjacent S-$p$ orbitals, and are found inside [above] the valence bands. The $e$-$pd$ and $a_{1}$-$pd$ states originate from the hybridization of the six S-$p$ and six (second nearest neighbour) Mo-$d$ orbitals and are situated in the gap.\cite{Noh.2014} We find that all presented states are stable and relaxed within our DFT framework. Since $\Delta E_{L}$ is significantly smaller than the quasi particle band gap in MoS$_{2}$, we assume that the Fermi energy is close to the conduction band minimum (CBM). We interpret the experimentally observed sharp emission lines as resulting from excitons that also involve particle-hole excitations from the defect state orbitals.\cite{refaely-abramson_defect-induced_2018}

Figures~\ref{fig1}f-i demonstrate that for instance, for $V^{2-}_{Mo}$ and $V^{3-}_{Mo}$, the DFT computed Fermi energy $E_{F}^{DFT}$ is consistently close to the CBM. Taking the computed energy differences between each state and the CBM, our ab-initio calculation suggests that in this case the lowest unoccupied state in the $V^{3-}_{Mo}$, the $e$-$pd$ state, is likely to be involved in the photoluminescence signal $X_{L}$ as in Fig.~\ref{fig1}d (cf. Supplementary Information). In particular, the computed energy difference $\Delta E^{DFT} = E(CBM)^{DFT} - E(e-pd)^{DFT} = \SI{0.22}{\electronvolt}$ coincides with the highest experimental detuning $\Delta E_{L}$ in the top panel of Fig.~\ref{fig1}d. However, the DFT calculation does not consider the binding energy of the exciton that is comprised of the different electronic orbitals with a complex distribution of weights and varying momentum dependencies.\cite{refaely-abramson_defect-induced_2018}


%
\begin{figure}
	\scalebox{\figurescale}{\includegraphics[width=1\linewidth]{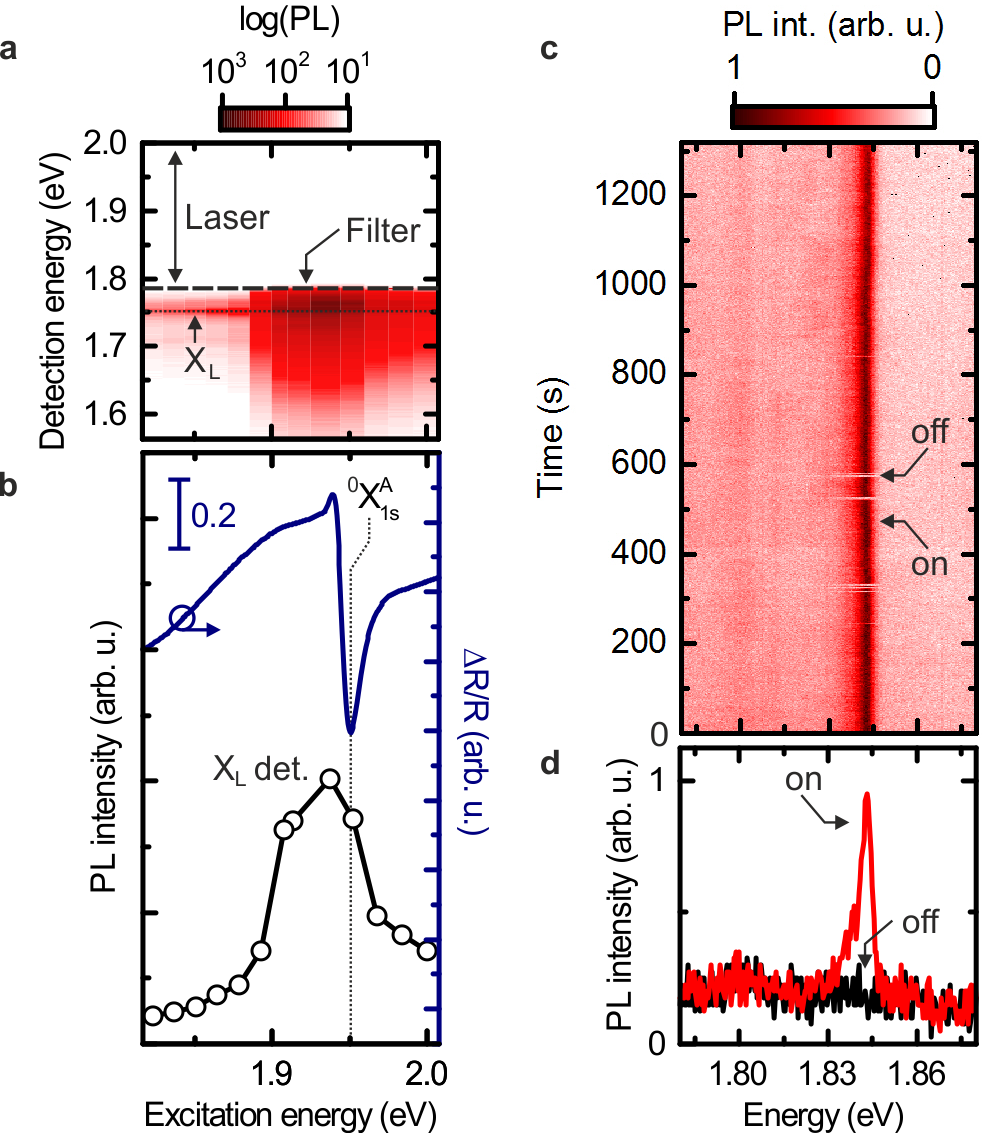}}
	\renewcommand{\figurename}{Fig.}
	\caption{\label{fig2}
		\textbf{PL excitation spectroscopy and time stability of optically active single defect emitters.}
		\textbf{a}, False color plot showing the localized emission $X_{L}$ for the excitation being tuned across $^{0}X^{A}_{1s}$.
		\textbf{b}, The differential reflectivity $\Delta R/R$ of the heterostructure reveals the $X^{A}_{1s}$ as highlighted by the dashed line. The reflectivity is compared to the PL intensity of the $X_{L}$ as a function of laser excitation energy.
		\textbf{c}, Time trace of the photoluminescence emission of a single defect emitter recorded over long times. The integration time is set to $\SI{1}{\second}$ while spectra are acquired every second. 
		\textbf{d}, Two exemplary spectra taken from \textbf{c} for the emitter being switched on and off.}
\end{figure}

To corroborate the above interpretation of a single optically active defect, we perform photoluminescence excitation (PLE) spectroscopy to probe the photo-physics and energetic structure of the involved states. Fig.~\ref{fig2}a presents a false color plot for a typical PLE scan of an emission line $X_{L}$ as the excitation energy is tuned across the $^{0}X^{A}_{1s}$. The emission occurs at $E(X_{L}) = \SI{1.76 \pm 0.01}{\electronvolt}$ energetically well below the $^{0}X^{A}_{1s}$. When measuring differential reflectivity $\Delta R/R$ using a broadband supercontinuum source, we identify $^{0}X^{A}_{1s}$ at an energy of $E(^{0}X^{A}_{1s}) = \SI{1.95}{\electronvolt}$ (cf. Fig.~\ref{fig2}b). Upon resonant excitation of $^{0}X^{A}_{1s}$, the $X_{L}$ emission is significantly increased. This is a consequence of the efficient creation of excitons, when the laser energy is on resonance with an exciton transition (cf. Supplementary Information for excitation on $^{0}X^{A}_{2s}$ and $^{0}X^{B}_{1s}$). Most significantly, we can excite $X_{L}$ states also below the free exciton transitions. This is consistent with recently predicted absorbance spectra of single vacancy defects in 2D materials \cite{refaely-abramson_defect-induced_2018} and in contrast to strain induced quantum dot-like emitters. \cite{Tonndorf.2015,Srivastava.2015,He.2015,Koperski.2015,Chakraborty.2015,Kern.2016,Branny.2017,PalaciosBerraquero.2017,PalaciosBerraquero.2017,Tonndorf.2017,Blauth.2018}

Another key property of quantum emitters is their spectral stability. A representative time trace of a single defect emitter is presented in Fig.~\ref{fig2}c (with a spectrum acquired every second). Two exemplary spectra highlighted in Fig.~\ref{fig2}c are plotted in Fig.~\ref{fig2}d. They demonstrate that the emitter exhibits a blinking behavior. This is an observed feature of quantum emitters and more generally of various types of few-level systems. \cite{Efros.2016} In our case, the blinking behavior is not surprising since the defects are embedded in a two-dimensional plane with large surface-to-volume ratio and charges in the environment that are very likely to randomly fluctuate in position and time.

\begin{figure}[t]
\scalebox{\figurescale}{\includegraphics[width=1\linewidth]{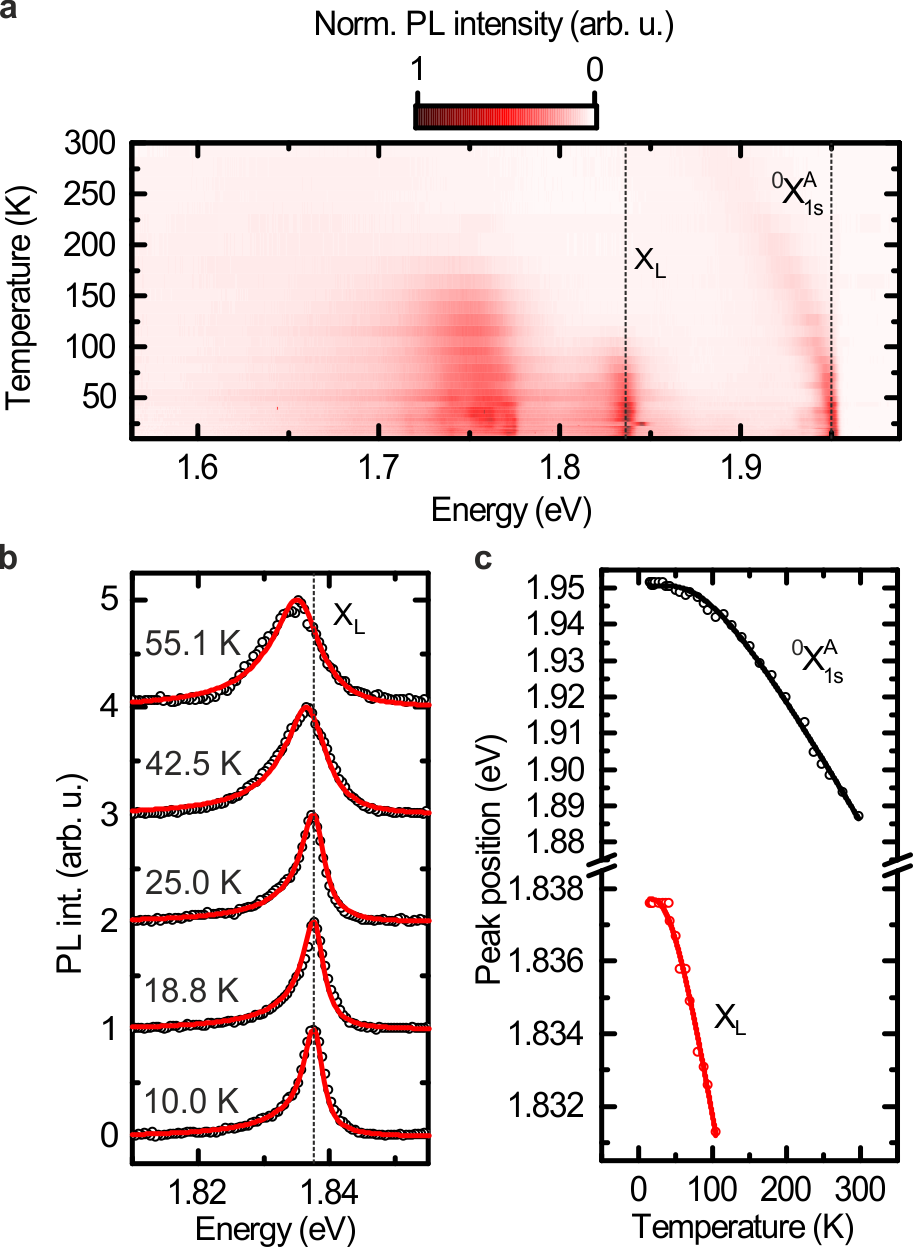}}
\renewcommand{\figurename}{Fig.}
\caption{\label{fig3}
\textbf{Temperature-dependent PL spectroscopy of a single defect and independent Boson model.} 
\textbf{a}, False color plot of the temperature-dependent evolution of PL from localized and delocalized excitons. The emission energies of neutral exciton $^{0}X^{A}_{1s}$ and one single defect emitter $X_{L}$ at $\SI{10}{\kelvin}$ are highlighted with a dashed line, respectively.
\textbf{b}, Temperature dependent spectra of $X_{L}$ fitted with an independent Boson model. Best agreement is found for $a_{B} = \SI{2}{\nano\meter}$.
\textbf{c}, Temperature dependent peak position of $^{0}X^{A}_{1s}$ and $X_{L}$. Data are fitted with Eq. \ref{eq:EnergyvsT}.
}
\end{figure}

Measuring the temperature dependence of the $\mu$-PL spectra allows us to determine the interactions between the involved electron orbitals and phonons. We gradually change the lattice temperature from $\SI{10}{\kelvin}$ to $\SI{300}{\kelvin}$ while recording $\mu$-PL spectra (cf. Fig.~\ref{fig3}a). The spectra reveal emission from the single defect emitters that are strongly redshifted and simultaneously broadened at elevated temperatures (similar as for the $^{0}X^{A}_{1s}$). 
We analyze the evolution of the emitter $X_{L}$ in detail by utilizing the independent Boson model that has been succcessfully applied to the lineshape analysis of quantum emitter states \cite{zimmermann_dephasing_2002,wilson-rae_quantum_2002,krummheuer_theory_2002} and defect-bound excitons.\cite{duke_phonon-broadened_1965} Figure~\ref{fig3}b highlights the spectra of $X_{L}$ for selected temperatures with lineshapes fitted by this model (red lines). In particular, the lineshape of each emitter is found to exhibit a very pronounced phonon sideband ($\sim \SI{10}{\milli\electronvolt}$) at the low energy side. Most likely, this low energy tail results from interaction with a specific bandwidth of acoustic phonons since the defects are highly localized in real space. \cite{refaely-abramson_defect-induced_2018} Our model particularly accounts for the coupling with LA/TA phonon branches, while coupling with the ZA branch is found to be negligible. For low temperatures, we find the best agreement of the calculated lineshape with our data for an effective Bohr radius of $a_{B} = \SI{2}{\nano\meter}$. Taking into account a phonon lifetime of $\frac{1}{\gamma_{ph}} \sim \SI{40}{\pico\second}$, \cite{gu_layer_2016} we obtain an excitonic radiative lifetime (linewidth) of $\sim\SI{1}{\pico\second}$ ($\sim \SI{0.5}{\milli\electronvolt}$). The lineshape in Fig.~\ref{fig3}b is most asymmetric at low temperatures where phonon emission is more likely to happen than phonon absorption while for higher temperatures, the lineshape becomes more symmetric.

%

The temperature dependent peak positions in Fig.~\ref{fig3}a manifest in the well-known polaron shift for all spectral features (compare Fig.~\ref{fig3}c), as phenomenologically described by \cite{ODonnell.1991}

\begin{equation}
E_{G}(T) = E_{G}(0)-S \langle \hbar \omega\rangle \bigl[ \coth \frac{\langle \hbar \omega\rangle }{ 2k_{B}T} -1 \bigr],
    \label{eq:EnergyvsT}
\end{equation}
with the emission energy $E_{G}(0)$ at $\SI{0}{\kelvin}$, the Huang-Rhys (HR) factor $S$, the average phonon energy $\langle \hbar \omega\rangle$, and the Boltzmann constant $k_{B}$. Fitting $X_{L}$ in Fig.~\ref{fig3}c reveals average phonon energies of $\langle \hbar \omega\rangle = \SI{12.69\pm1.17}{\milli\electronvolt}$ and an electron-phonon coupling (HR factor) of $S = 0.791 \pm 0.071$ (for $^{0}X_{1s}^{A}$: $\langle \hbar \omega\rangle = \SI{25.11\pm1.02}{\milli\electronvolt}$ and $S = 2.135 \pm 0.061$). The experimental HR factor is in very good agreement with a HR factor of $0.75$ as obtained from the independent Boson model (cf. Supplementary Information).

\section{Discussion}

The following arguments underline that most likely HIM induced Mo-vacancies give rise to the sharp emission lines. Importantly, the emission peaks occur homogeneously across the whole basal plane of the bombarded MoS$_{2}$ encapsulated into hBN (cf. Fig.~\ref{fig1}c). The sharp peaks do not occur in bombarded hBN without MoS$_{2}$ nor in non-bombarded hBN/MoS$_{2}$/hBN (cf. Supplementary Information). This phenomenology strongly suggests that the optically active defects can not be explained by single S-vacancies, because the latter are ubiquitous in exfoliated, non-bombarded MoS$_{2}$ single-layers \cite{Hong.2015} although they are very likely to be produced under helium ion exposure as well. \cite{Komsa.2012,Noh.2014,Klein.2017} However, S-vacancies suffer from passivation by oxygen,\cite{barja2018} which especially applies to the discussed nano-fabricated TMDCs under ambient conditions. Equally, quantum emitters in hBN can be excluded (cf. Supplementary Information).\cite{Tran.2015} Intriguingly, the beam energy of the HIM ($\SI{30}{\kilo\electronvolt}$) and also the excess energy of the secondary electrons (< $\SI{10}{\electronvolt}$) produced in the exposed materials are sufficient to form Mo-vacancies in the MoS$_{2}$, \cite{Ohya.2009,Noh.2014} that are rare in pristine samples.\cite{Hong.2015} This argument could explain why we observe optically active defect sites only after the helium ion exposure of the MoS$_{2}$ monolayers.

In our understanding, the influence of hBN is two-fold. On the one hand, the emission peaks $X_{L}$ have the lowest emission energy of $\Delta E_{L}$ of $\sim \SI{0.2}{\electronvolt}$ below $^{0}X_{1s}^{A}$. Encapsulated in hBN, the excitonic emission energy is renormalized due to dielectric environment, \cite{Florian.2017} and particularly, the encapsulation shifts the emission energy closer to the $^{0}X_{1s}^{A}$ due to a decrease of the local Coulomb interactions. On the other hand, the emission linewidth is strongly reduced for both $X_{L}$ and $^{0}X_{1s}^{A}$. \cite{Wierzbowski.2017,Ajayi.2017,Cadiz.2017,Florian.2017} The combination of the hBN-effects explain the sharp spectral features which are spread across the broad energy window $\Delta E_{L} \sim \SI{100}{\milli\electronvolt} - \SI{220}{\milli\electronvolt}$ in Fig.~\ref{fig1}d. For all possibly created defects, the emission at $\Delta E_{L}$ of $\sim \SI{0.2}{\electronvolt}$ must resemble the lowest unoccupied, optically active state. This interpretation is consistent with the saturating behavior as in Fig.~\ref{fig1}e, and it also explains the blinking behavior in Fig.~\ref{fig2}c-d. Remarkably, the PLE measurements in Fig.~\ref{fig2}a are also consistent with this interpretation, since the expected joint density of states of localized electron orbitals at the defect and its environment sum up to a finite, continuous absorbance from the lowest unoccupied state (i.e. at $\Delta E_{L}$ of $\sim \SI{0.2}{\electronvolt}$) up to $^{0}X_{1s}^{A}$. \cite{refaely-abramson_defect-induced_2018} The excitation laser creates electrons that can occupy the defect states followed by optical recombination with a valence band hole and possible further local defect orbitals.\cite{refaely-abramson_defect-induced_2018} In other words, the complex admixture of $p$- and $d$-orbitals in the Mo-defect (compare Fig.~\ref{fig1}f-i) give rise to a continuous absorbance below $^{0}X_{1s}^{A}$, as recently discussed for single chalcogen vacancies. \cite{Noh.2014,refaely-abramson_defect-induced_2018} Our DFT calculation neglects the momentum dependence of the involved electron states and variable binding energies per orbital. \cite{refaely-abramson_defect-induced_2018} In the experiment, the influence of ambient gases might have further effects.\cite{Klein.2017}

Fig.~\ref{fig3}b demonstrates that the effective exciton-phonon interaction can be described by the independent boson model, and the quality of the fits suggests that the effective interaction length is localized within $\sim \SI{2}{\nano\meter}$. The independent boson model further suggests an excitonic lifetime of only 1 ps (cf. Supplementary Information). This theoretical value is a lower bound for the actual lifetime because we omit other sources of broadening mechanisms. Experimentally, we determine the upper bound of the radiative lifetime to be <$\SI{150}{\pico\second}$ (limited by the instrument response function), which renders second order time-correlation measurements of g$^{(2)}$ challenging. This experimental observation is in stark contrast to strain induced quantum dot-like emitters in TMDCs with nanosecond lifetimes based on an effective type-II band alignment.\cite{Tonndorf.2015,Srivastava.2015,He.2015,Koperski.2015,Chakraborty.2015} Future gated devices may allow us to reduce the background PL of the so-called L-peak, so that the quantum nature of the emitted light can be further analyzed.







\section{Conclusion}

In conclusion, we demonstrate a methodology to deterministically generate optically active single defects in MoS$_{2}$. The superior optical quality resulting from hBN encapsulation reveals spectrally narrow emission at very distinct detunings with respect to the neutral exciton. An observed saturating power dependence and a blinking behavior are consistent with the creation of single quantum emitters. On a broader perspective, a controlled production of optically active defects in a periodic pattern using helium ion beam bombardment may allow the exploration of exotic many-body physics in lattice systems, including the demonstration of coupled photon sources,~\cite{aharonovich_solid-state_2016} the Anderson orthogonality catastrophe,~\cite{anderson_infrared_1967, schmidt_Universal_2018} as well as the Mott transition between a superfluid and an insulator~\cite{bloch_many-body_2008} in the presence of dissipation. Moreover, the inherent proximity of these engineered defect states to surfaces and their nanometer size opens exciting ways for harnessing them as quantum sensors, investigating nuclear spin physics with wave functions only sampling a fraction of atoms or studying screening physics in the ultimate limit of a single localized electronic defect state.

\newpage

\section{Methods}
\subsection{Sample structure}
We employed the viscoelastic transfer method to transfer MoS$_2$ single-layer crystals and hBN multilayers onto $\SI{290}{\nano\meter}$ SiO$_2$ substrates. The MoS$_{2}$ crystals used for exfoliaten were purchased from SPI Supplies while high-quality hBN bulk crystals were provided by Takashi Taniguchi and Kenji Watanabe from NIMS, Japan.

\subsection{Helium Ion Microscopy}
We used MoS$_2$ monolayers transfered onto hBN/SiO$_2$/Si substrates for He$^{+}$ irradiation. A beam current $I \sim \SI{1}{\pico\ampere}$ and a beam energy of $\SI{30}{\kilo\volt}$ are used. Large areas are exposed with a beam spacing of $\SI{5}{\nano\meter}$. The dwell time was adjusted such that a dose of $\sim 10^{12}$ He$^{+}$ $\SI{}{\per\centi\meter\squared}$ is obtained which is optimized for a high density of single defect emitters (cf. Supplementary Information).

\subsection{Optical spectroscopy}

Confocal optical spectroscopy was performed in a helium flow cryostat with the sample kept at a lattice temperature of $\sim \SI{10}{\kelvin}$.  For cw $\mu$-PL experiments, we used a doubled Nd:YAG laser with an excitation energy of $\SI{2.33}{\electronvolt}$ and typical excitation powers of less than $\SI{10}{\micro\watt}$ which results in an excitation power density of $\SI{0.884}{\kilo\watt\per\centi\meter\squared}$ at a laser spot diameter of $\SI{1.2}{\micro\meter}$. \\

For measuring differential reflectivity spectra a supercontinuum white light source was used that was focused with an optical microscope onto the sample kept at a lattice temperature of $\sim \SI{10}{\kelvin}$ with a spot size of $\sim 1- \SI{2}{\micro\meter}$.\\

For PLE experiments, we used an optical parametric oscillator (OPO) pumped with a Ti:Sa at $\SI{820}{\nano\meter}$ with a repetition rate of $\SI{80}{\mega\hertz}$ and a $\SI{150}{\femto\second}$ pulse width. For our measurements, we used the signal output of the OPO which was incrementally tuned in steps of $\sim \SI{25}{\milli\electronvolt}$ from $\SI{1.8}{\electronvolt}-\SI{2.45}{\electronvolt}$. The excitation power was kept at $\SI{20}{\micro\watt}$ for all excitation energies. The excitation laser pulse was filtered with a sharp edge filter before being dispersed on a grating and detected with a charge-coupled device (CCD).

\subsection{DFT calculations}
The calculations were performed using density functional theory (DFT) as implemented in the Vienna ab initio simulation package (VASP)\,\cite{VASP:3,VASP:4}. The projected augmented wave method has been used \cite{Kresse:99,Bloechl}. The atomic and electronic structures were determined using the PBE functional. A plane wave basis with an energy cutoff of ${\rm E_{cut}=500\,eV}$ and a $(6\times6\times1)$ Monkhorst-Pack {\bf k}-point sampling has been used. The TMD layer has been modeled using a $(9\times9)$ supercell containing 242 atoms.

%
%
\section{Acknowledgements}
Supported by Deutsche Forschungsgemeinschaft (DFG) through the TUM International Graduate School of Science and Engineering (IGSSE). We gratefully acknowledge financial support of the German Excellence Initiative via the Nanosystems Initiative Munich and the PhD program ExQM of the Elite Network of Bavaria. We also gratefully acknowledge financial support from the European Union’s Horizon 2020 research and innovation programme under grant agreement No. 820423 (S2QUIP) the German Federal Ministry of Education and Research via the funding program Photonics Research Germany (contract number 13N14846) and the Bavarian Academy of Sciences and Humanities. M.L. and M.F. were supported by the Deutsche Forschungsgemeinschaft (DFG) within RTG 2247 and through a grant for CPU time at the HLRN (Hannover/Berlin). J.C. is supported by NSF-DMR1410599 and the Visiting Professor Program from the Bavarian State Ministry for Science, Research \& the Arts. M.Kn. acknowledges support from the Technical University of Munich - Institute for Advanced Study, funded by the German Excellence Initiative and the European Union FP7 under grant agreement 291763 and the German Excellence Strategy Munich Center for Quantum Science and Technology (MCQST). R.S. acknowledges support from the Munich Center for Quantum Science and Technology (MCQST).

\section{Author contributions}
J.K., U.W., M.Ka., J.J.F. and A.W.H. conceived and designed the experiments, M.L. and M.F. performed the DFT and independent Boson model calculations, F.S. and J.K. prepared the samples, K.W. and T.T. provided high-quality hBN bulk crystals, J.K. performed helium ion exposure of samples, J.K., J.W., J.C. and K.M. performed the optical measurements, J.K. analyzed the data, R.S. and M.Kn. contributed interpreting the data, J.K. and A.W.H. wrote the manuscript with input from all coauthors.

%
\section{Additional information}

\subsection{Supplementary Information} accompanies this paper
\subsection{Competing financial interests} The authors declare no competing financial interests.

%
%
\bibliographystyle{achemso}
\bibliography{full}

\end{document}


\title{Atomistic defect states as quantum emitters in monolayer MoS$_2$}
%
\author{J.~Klein}\email{julian.klein@wsi.tum.de}
\affiliation{Walter Schottky Institut and Physik Department, Technische Universit\"at M\"unchen, Am Coulombwall 4, 85748 Garching, Germany}
\affiliation{Nanosystems Initiative Munich (NIM), Schellingstr. 4, 80799 Munich, Germany}
\author{M.~Lorke}
\affiliation{Institut für Theoretische Physik, Universität Bremen, P.O. Box 330 440, 28334 Bremen, Germany}
\author{M.~Florian}
\affiliation{Institut für Theoretische Physik, Universität Bremen, P.O. Box 330 440, 28334 Bremen, Germany}
\author{F.~Sigger}
\affiliation{Walter Schottky Institut and Physik Department, Technische Universit\"at M\"unchen, Am Coulombwall 4, 85748 Garching, Germany}
\affiliation{Nanosystems Initiative Munich (NIM), Schellingstr. 4, 80799 Munich, Germany}
\author{J.~Wierzbowski}
\affiliation{Walter Schottky Institut and Physik Department, Technische Universit\"at M\"unchen, Am Coulombwall 4, 85748 Garching, Germany}
\author{J.~Cerne}
\affiliation{Department of Physics, University at Buffalo, The State University of New York, Buffalo, New York 14260, USA}
\author{K.~M\"uller}
\affiliation{Walter Schottky Institut and Physik Department, Technische Universit\"at M\"unchen, Am Coulombwall 4, 85748 Garching, Germany}
\author{T.~Taniguchi}
\affiliation{National Institute for Materials Science, Tsukuba, Ibaraki 305-0044, Japan}
\author{K.~Watanabe}
\affiliation{National Institute for Materials Science, Tsukuba, Ibaraki 305-0044, Japan}
\author{U.~Wurstbauer}
\affiliation{Walter Schottky Institut and Physik Department, Technische Universit\"at M\"unchen, Am Coulombwall 4, 85748 Garching, Germany}
\affiliation{Nanosystems Initiative Munich (NIM), Schellingstr. 4, 80799 Munich, Germany}
\author{M.~Kaniber}
\affiliation{Walter Schottky Institut and Physik Department, Technische Universit\"at M\"unchen, Am Coulombwall 4, 85748 Garching, Germany}
\affiliation{Nanosystems Initiative Munich (NIM), Schellingstr. 4, 80799 Munich, Germany}
\author{M.~Knap}
\affiliation{Department of Physics and Institute for Advanced Study, Technical University of Munich, 85748 Garching, Germany}
\author{R.~Schmidt}
\affiliation{Max-Planck-Institut f\"ur Quantenoptik, 85748 Garching, Germany}
\author{J.~J.~Finley}\email{finley@wsi.tum.de}
\affiliation{Walter Schottky Institut and Physik Department, Technische Universit\"at M\"unchen, Am Coulombwall 4, 85748 Garching, Germany}
\affiliation{Nanosystems Initiative Munich (NIM), Schellingstr. 4, 80799 Munich, Germany}
\author{A.~W.~Holleitner}\email{holleitner@wsi.tum.de}
\affiliation{Walter Schottky Institut and Physik Department, Technische Universit\"at M\"unchen, Am Coulombwall 4, 85748 Garching, Germany}
\affiliation{Nanosystems Initiative Munich (NIM), Schellingstr. 4, 80799 Munich, Germany}
%

%
\date{\today}

\maketitle
%
%

\tableofcontents

\renewcommand{\theequation}{S\arabic{equation}}

\newpage

\section{Optical spectroscopy of He-ion bombarded hBN}

%
%
\begin{figure}[!ht]
\scalebox{\figurescale}{\includegraphics[width=1.2\linewidth]{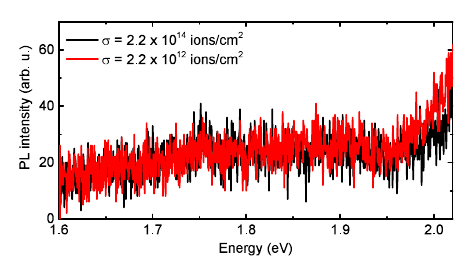}}
\renewcommand{\figurename}{SI Fig.}
\caption{\label{SIfig3}
\textbf{Photoluminescence of He-ion bombarded hBN.}
Two exemplary spectra taken from a spatially resolved low-temperature ($\SI{10}{\kelvin}$) $\mu$-PL mapping on He-ion bombarded hBN with doses of $\sigma = 2.2 \cdot 10^{12}$ ions $\SI{}{\per\centi\meter\squared}$ (red) and $\sigma = 2.2 \cdot 10^{14}$ ions $\SI{}{\per\centi\meter\squared}$ (black).
}
\end{figure}

Recently, quantum emission from optically active defects in hBN crystals has been demonstrated. \cite{Tran.2015} In order to rule out any defect emission contribution to the defect emission from the hBN crystals, we irradiate large areas (>$\SI{400}{\micro\meter\squared}$) of exfoliated and deterministically transferred hBN flakes on the same SiO$_{2}$/Si substrates that are used for all experiments of this work. Two spectra taken from low temperature ($\SI{10}{\kelvin}$) spatially resolved $\mu$-PL mappings recorded for He-ion doses of $\sigma = 2.2 \cdot 10^{12}$ ions $\SI{}{\per\centi\meter\squared}$ (red) and $\sigma = 2.2 \cdot 10^{14}$ ions $\SI{}{\per\centi\meter\squared}$ (black) are shown in SI Fig.~\ref{SIfig3}. Besides background noise from the CCD chip used for integration in our experiments, our data shows no indication of quantum emission from hBN induced by the He-ions in all bombarded areas.

\newpage

\section{Dose dependent spectra of non-encapsulated MoS$_{2}$}

%
%
\begin{figure}[!ht]
\scalebox{\figurescale}{\includegraphics[width=1.2\linewidth]{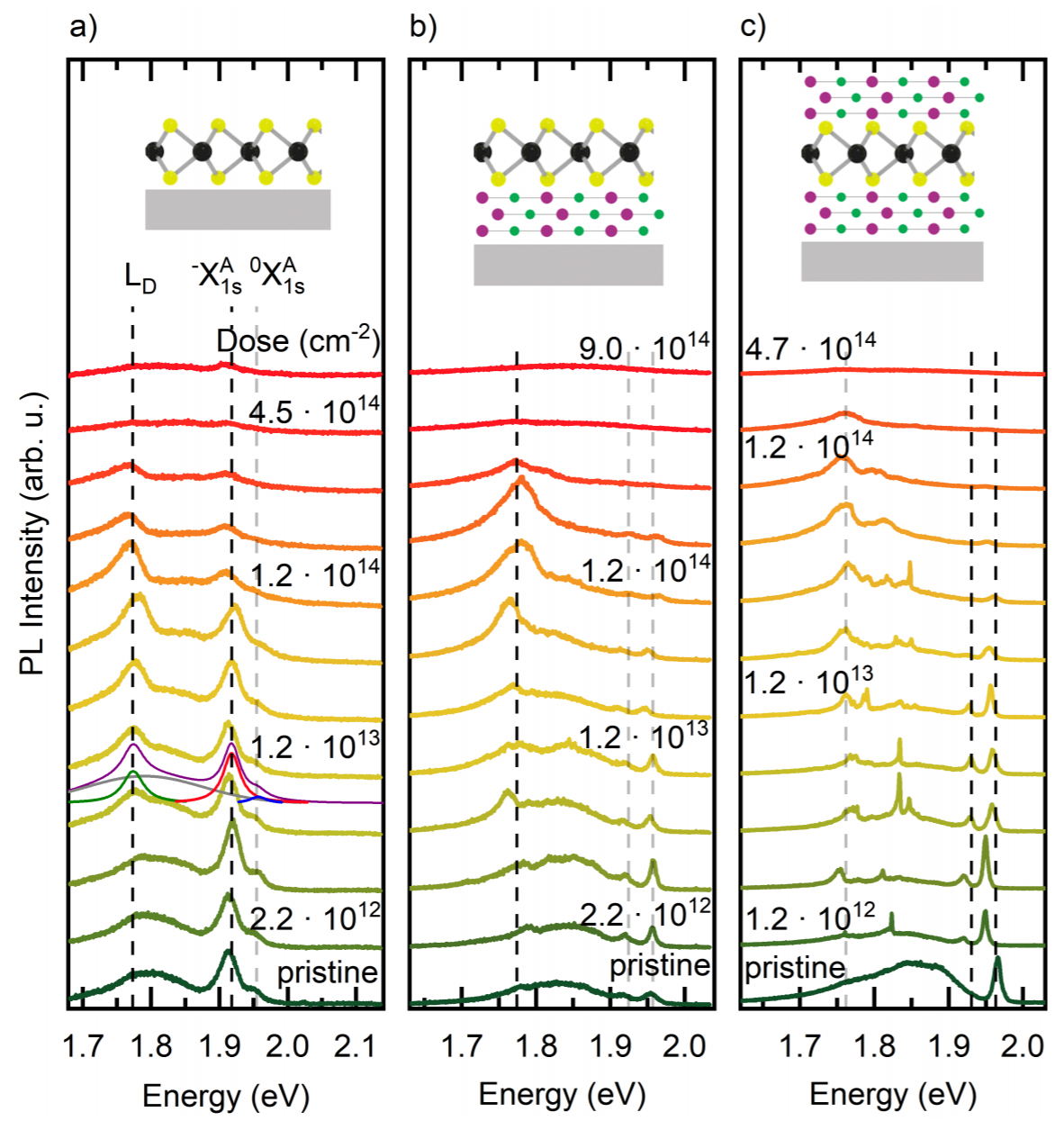}}
\renewcommand{\figurename}{SI Fig.}
\caption{\label{SIfig4}
\textbf{He-ion dose dependent photoluminescence spectra of MoS$_{2}$ in different dielectric environments.}
\textbf{a}, Dose dependent photoluminescence of MoS$_{2}$ on SiO$_{2}$/Si. The spectrum reveals emission from the neutral exciton $^{0}X^{A}_{1s}$, and charged exciton $^{-}X^{A}_{1s}$, and from the L-peak and furthermore dose dependent emission from the L$_{D}$ peak. (Data are adapted from Ref. \cite{Klein.2017})
\textbf{b}, Dose dependent photoluminescence of MoS$_{2}$ on hBN reveals spectrally more narrow free exciton emission and similar to \textbf{a} emission from the L-peak and dose dependent defect emission from the L$_{D}$-peak
\textbf{c}, Dose dependent photoluminescence of a hBN/MoS$_{2}$/hBN van der Waals heterostructure. Besides spectrally narrow free exciton emission the spectrum also reveals spectrally sharp emission.}
\end{figure}

Since the photo-physical properties of single-layer MoS$_{2}$ strongly depend on the dielectric environment, especially on the encapsulation and passivation of the surface of the crystal,\cite{Cadiz.2016,Wierzbowski.2017,Cadiz.2017,Florian.2017} we spectroscopically investigate the He-ion dose dependence of single-layer MoS$_{2}$ in different dielectric environments. SI Fig.~\ref{SIfig4} shows typical low-temperature ($\SI{10}{\kelvin}$) $\mu$-PL spectra of single-layer MoS$_{2}$ on SiO$_{2}$/Si, MoS$_{2}$/hBN and fully hBN encapsulated MoS$_{2}$. The He-ion dose is varied between $\sigma \sim 10^{12}$ ions $\SI{}{\per\centi\meter\squared}$ and $\sigma \sim 10^{15}$ ions $\SI{}{\per\centi\meter\squared}$. All spectra reveal emission from the neutral and charged exciton with much narrower linewidths on hBN substrates due to a reduced inhomogeneous linewidth. \cite{Wierzbowski.2017,Ajayi.2017,Cadiz.2017,Florian.2017} Moreover, all spectra exhibit typically observed broad low-energy emission from the L-peak in addition to a superimposed emission from the L$_{D}$ peak \cite{Klein.2017} that increases with increased $\sigma$. This emission is detuned by $\Delta E \sim \SI{190}{\milli\electronvolt}$ from the neutral exciton and spectrally narrows for fully hBN encapsulated MoS$_{2}$. Remarkably, only sandwiched MoS$_{2}$ reveals very sharp emission. The overall PL emission in all three sample geometries quenches for $\sigma > 10^{14}$ ions $\SI{}{\per\centi\meter\squared}$ which is likely due to high defect densities.

\newpage

\section{Spatially localized defect emission in vdW heterostructures}

%
%
\begin{figure}[!ht]
\scalebox{\figurescale}{\includegraphics[width=1.7\linewidth]{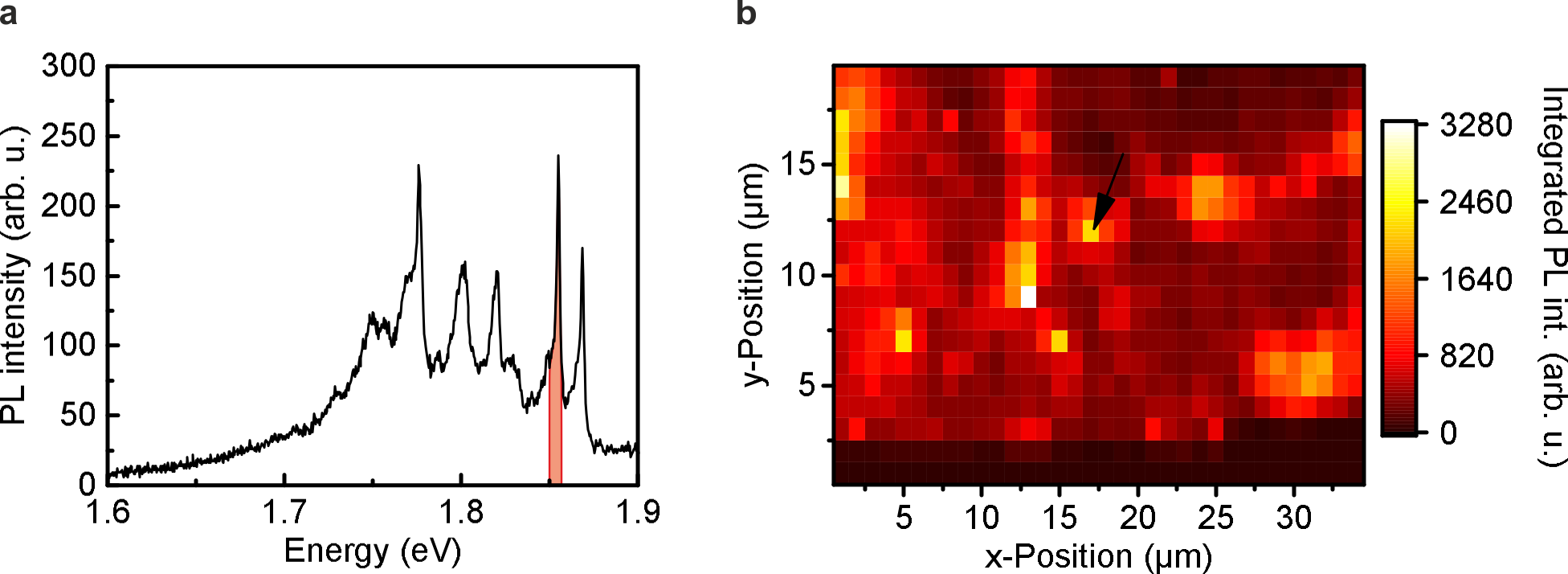}}
\renewcommand{\figurename}{SI Fig.}
\caption{\label{SIfig5}
\textbf{Spatially localized photoluminescence in hBN/MoS$_{2}$/hBN.}
\textbf{a}, Typical photoluminescence spectrum that features emission from different single defect emitters. 
\textbf{b}, Corresponding spatially integrated PL mapping of the red coloured line in \textbf{a}. The spatial position from which the spectrum is taken from is highlighted by the black arrow. The photoluminescence is spatially localized within the focal spot diameter of the confocal microscope.
}
\end{figure}

To spatially correlate the spectrally sharp emission, we perform spatially resolved $\mu$-PL mappings. SI Fig.~\ref{SIfig5}a shows a typical photoluminescence spectrum at $\SI{10}{\kelvin}$. The spectrum reveals various single defect emitters. The corresponding spatially integrated PL mapping for the red coloured emitter is shown in SI Fig.~\ref{SIfig5}b. The arrow highlights the spatial position from which the spectrum is taken from. The emission is spatially localized with a spatial extent that is limited by the focal spot diameter ($\sim \SI{1.2}{\micro\meter}$) of our confocal microscope.

\newpage

\section{Averaged photoluminescence spectrum of He-ion bombarded hBN/MoS$_{2}$/hBN}

%
%
\begin{figure}[!ht]
\scalebox{\figurescale}{\includegraphics[width=1.4\linewidth]{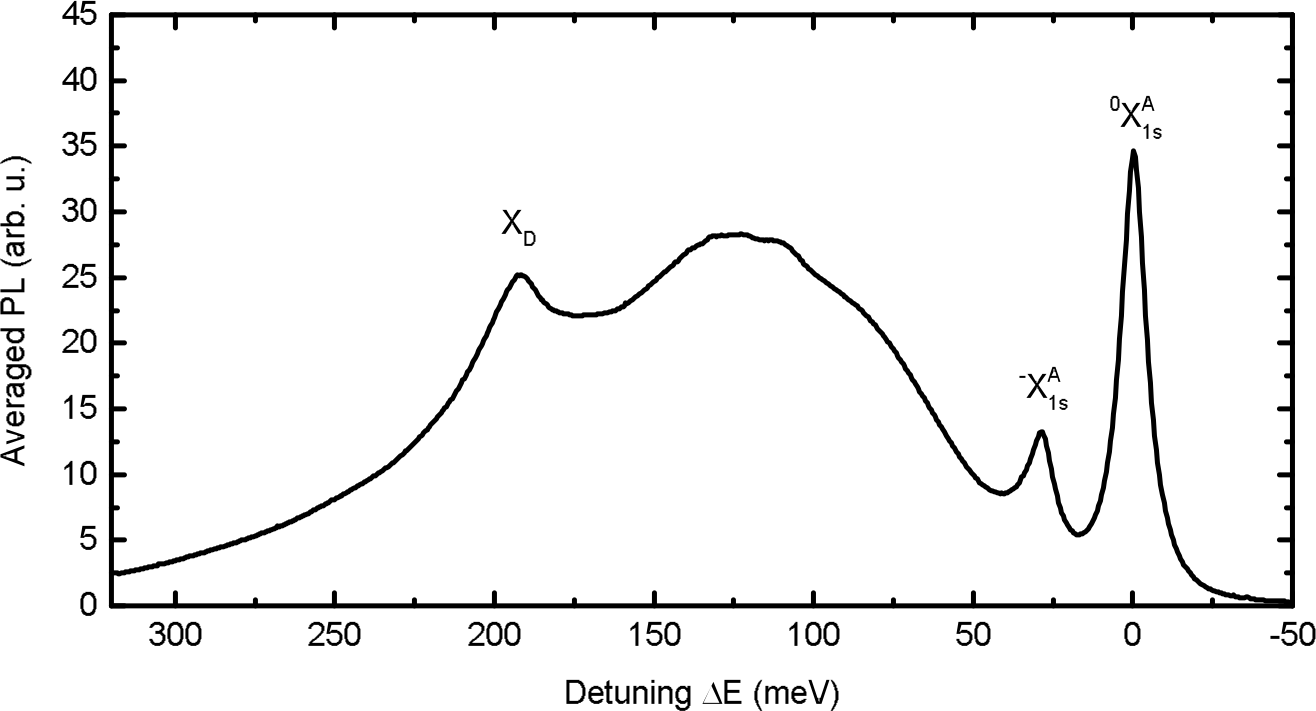}}
\renewcommand{\figurename}{SI Fig.}
\caption{\label{SIfig6}
\textbf{Averaged photoluminescence spectrum of defective, and hBN encapsulated MoS$_{2}$.}
The spatially averaged photoluminescence spectrum shows emission from neutral and charged exciton and also significant contribution from the L-peak and from the defect X$_{D}$-peak. The neutral exciton is set to zero detuning $\Delta E$.
}
\end{figure}

SI Fig.~\ref{SIfig6}a shows an averaged PL spectrum with the neutral exciton set to zero detuning $\Delta E$. Here, we average over all spectra that reveal emission from the neutral exciton and emission from localized emitters. In each spectrum the emission energy of the neutral exciton is used as a reference in order to properly sum up all spectra. The averaged spectrum shows emission from the neutral and charged exciton, a significant contribution from the L-peak at lower energies and also a significant contribution from the most prominent defect peak X$_{D}$ at $\Delta E \sim \SI{190}{\milli\electronvolt}$. The spectrum has similarities with a photoluminescence spectrum of He-ion bombarded MoS$_{2}$ on SiO$_{2}$ as reported recently~\cite{Klein.2017} and also shown in SI Fig.~\ref{SIfig4}a. 


\newpage

\section{Spectral occurence of single defect emission}

%
%
\begin{figure}[!ht]
\scalebox{\figurescale}{\includegraphics[width=0.8\linewidth]{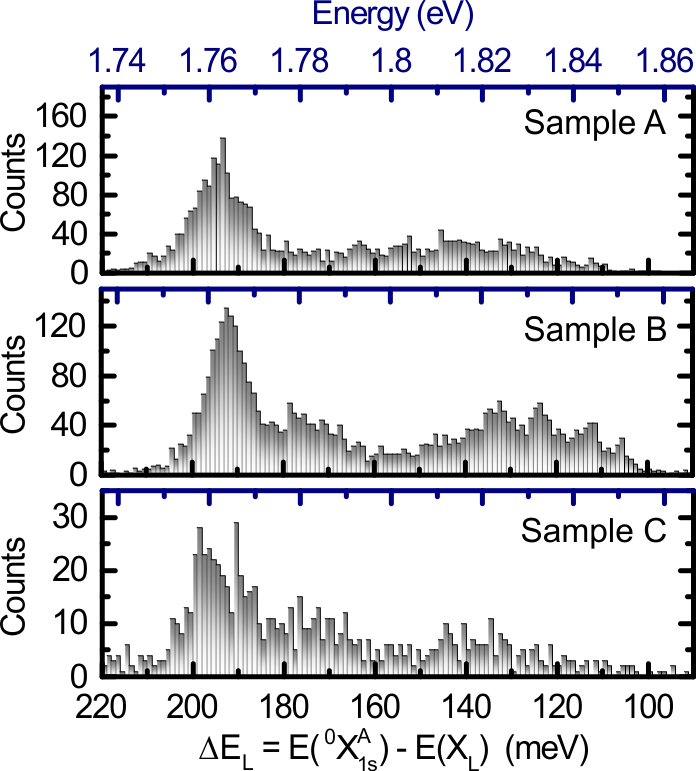}}
\renewcommand{\figurename}{SI Fig.}
\caption{\label{SIfigstatistics}
\textbf{Photoluminescence of He-ion bombarded hBN.}
Histogram of the energy detuning $\Delta E_{L}$ of single defect emission $X_{L}$ for three samples. Data are obtained from low-temperature ($\SI{10}{\kelvin}$) spatially resolved $\mu$-PL spectroscopy.
}
\end{figure}
To verify the reproducability of spectral occurence of single defect emission, we repeated experiments on two additional helium ion bombarded heterostacks. To this end, perform low-temperature ($\SI{10}{\kelvin}$) spatially resolved $\mu$-PL spectroscopy on three nominally identical heterostacks. Emission statistics for all three measured samples are presented in SI Fig.~\ref{SIfigstatistics}. All histograms qualitatively reveal the same distribution, showing clustering at $\Delta E \sim \SI{190}{\milli\electronvolt}$ and a broader emission band for $< \Delta E \sim \SI{190}{\milli\electronvolt}$.

\newpage

\section{Dose dependent occurence of single defect emitters}

%
%
\begin{figure}[!ht]
\scalebox{\figurescale}{\includegraphics[width=0.8\linewidth]{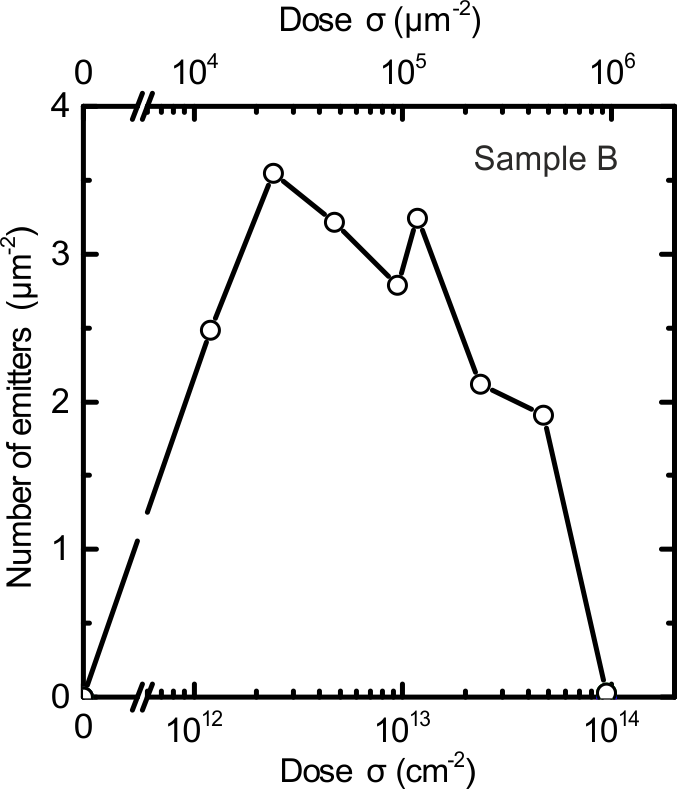}}
\renewcommand{\figurename}{SI Fig.}
\caption{\label{SIfig7}
\textbf{Dose dependent occurence of single defect emission.}
The dose dependent number of emitters shows a maximum at a dose of $\sigma \sim 2 \cdot 10^{12} \SI{}{\per\centi\meter\squared}$ with a steep decrease for $\sigma > 7 \cdot 10^{13} \SI{}{\per\centi\meter\squared}$.
}
\end{figure}

Instead of creating defects with a fixed helium ion dose, we continuously vary the ion dose over two orders of magnitude from $\sigma = 2.2 \cdot 10^{12} \SI{}{\per\centi\meter\squared}$ to $\sigma = 1.4 \cdot 10^{14} \SI{}{\per\centi\meter\squared}$ by adjusting the dwell time accordingly. Here, large fields of $4 \times \SI{8}{\micro\meter}$ are exposed for each dose in order to create reliable statistics. The creation efficiency of single defect emitters has a maximum at $\sim 2 \cdot 10^{12} \SI{}{\per\centi\meter\squared}$ corresponding to $\sim 3.5$ emitters $\SI{}{\per\micro\meter\squared}$. 


\newpage

\section{Photoluminescence excitation spectroscopy}

%
\begin{figure}[!ht]
\scalebox{\figurescale}{\includegraphics[width=0.7\linewidth]{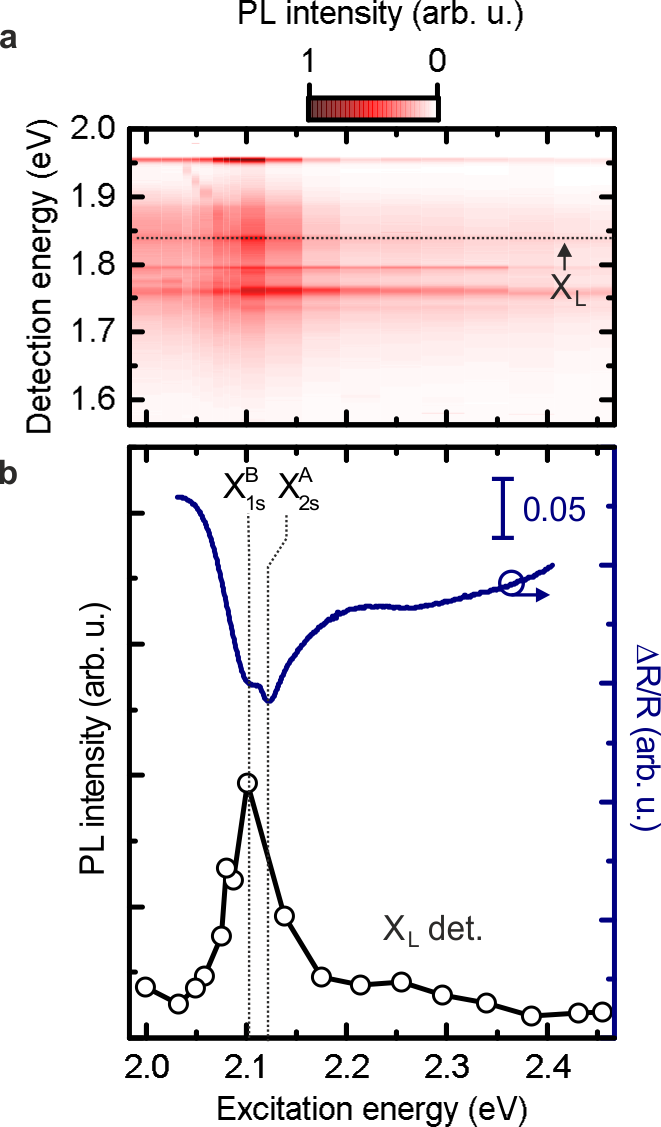}}
\renewcommand{\figurename}{SI Fig.}
\caption{\label{figPLE}
\textbf{Photoluminescence excitation spectroscopy of single defect emitters in a helium ion bombarded hBN/MoS$_{2}$/hBN van der Waals heterostructure.} 
\textbf{a}, The false colour plot shows emission from $^{0}X^{A}_{1s}$ and localized emission $X_{L}$ for excitation energetically above $^{0}X^{A}_{1s}$. 
\textbf{b}, Differential reflectivity $\Delta R/R$ of the heterostructure reveals the $X^{A}_{2s}$ and $X^{B}_{1s}$ as highlighted by the dashed lines. Photoluminescence intensity of $X_{L}$ as a function of the laser excitation energy. The intensity enhancement of single defect emission and background coincides well when the laser is tuned on resonance with $X^{A}_{2s}$ and $X^{B}_{1s}$.
}
\end{figure}

Similar to the PLE measurements shown in Fig.~2. in the main manuscript, we performed additional measurements by scanning our excitation laser across excitonic resonances energetically above the $^{0}X^{A}_{1s}$ (cf. SI Fig.~\ref{figPLE}a). By measuring differential reflectivity $\Delta R/R$ using a broadband supercontinuum source, we observe several excitonic resonances as shown in SI Fig.~\ref{figPLE}b. The plot reveals absorption from the spin-orbit split $^{0}X^{B}_{1s}$ at an energy $E(^{0}X^{B}_{1s}) = \SI{2.1}{\electronvolt}$ and an additional resonance close to the $^{0}X^{B}_{1s}$ at $E(^{0}X^{A}_{2s}) = \SI{2.12}{\electronvolt}$ which is ascribed to originate from the $^{0}X^{A}_{2s}$ as identified recently.\cite{Robert.2018} The hBN environment enhances optical quality and therefore allows us to observe higher excited Rydberg states.

\newpage

\section{He-ion exposed hBN prior to stacking of MoS$_{2}$}

%
%
\begin{figure}[!ht]
\scalebox{\figurescale}{\includegraphics[width=1.7\linewidth]{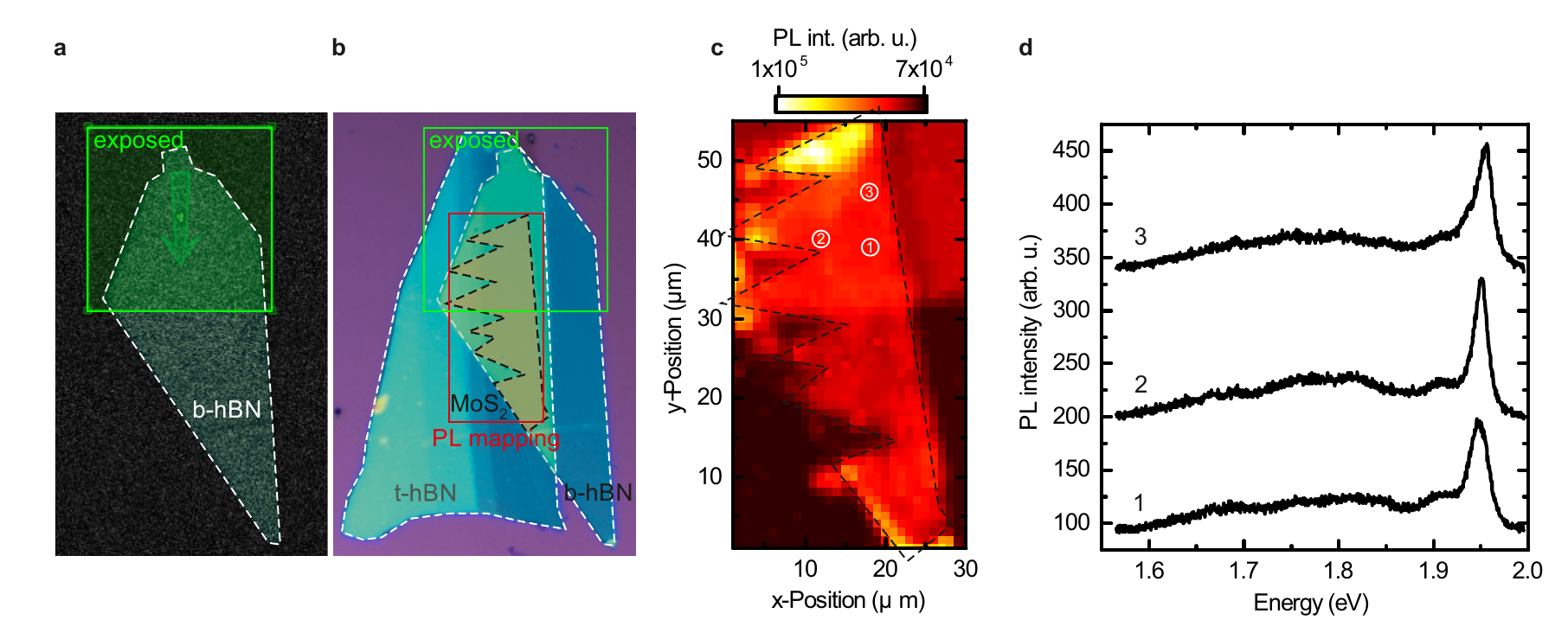}}
\renewcommand{\figurename}{SI Fig.}
\caption{\label{SIfig12}
\textbf{Spatially resolved photoluminescence of bombarded hBN with pristine MoS$_{2}$ and hBN stacked on top.}
\textbf{a}, HIM image of the bottom hBN flake. The highlighted He-ion exposed area is treated with a dose of $\sigma = 2.2 \cdot 10^{12} \SI{}{\per\centi\meter\squared}$.
\textbf{b}, Optical microscope image of the hBN/MoS$_{2}$/hBN van der Waals heterostructure.
\textbf{c}, Spatially resolved and spectrally integrated ($1.6-\SI{2.0}{\electronvolt}$) $\mu$-PL mapping of the highlighted area in \textbf{b}.
\textbf{d}, Representative spectra of three different positions taken from the PL mapping in \textbf{c}.
}
\end{figure}

In order to rule out any correlation of exposure induced photoluminesce with defects created in hBN through He-ion exposure we investigate van der Waals heterostructures where the bottom hBN flake is treated with He-ions. SI Fig.~\ref{SIfig12}a depicts a HIM image of a hBN flake on a SiO$_{2}$ substrate. The highlighted area of the flake is exposed with He-ions with a dose of $\sigma = 2.2 \cdot 10^{12} \SI{}{\per\centi\meter\squared}$. This dose has shown to result in a high density of single defect emitters (cf. SI Fig~\ref{SIfig7}). After the bombardment a single layer of MoS$_{2}$ is stacked onto the partially bombareded b-hBN and is in a subsequent step fully enapsulated with the t-hBN crystal. A spatially resolved and spectrally integrated $\mu$-PL mapping of the highlighted area in fig \ref{SIfig12}b is presented in SI Fig~\ref{SIfig12}c. Representative spectra from three different positions as highlighted in SI Fig~\ref{SIfig12}c are shown in SI Fig~\ref{SIfig12}d. Here, the spectra reveal luminescence from the neutral and charged exciton as well as the L-peak, signified by the broad luminescence background at lower energies. Importantly, unlike bombarded and encapsulated MoS$_{2}$ as presented in the main manuscript, the spectra from bombarded hBN do not feature any single defect emitters. We therefore can unambiguously exclude He-bombardment induced defects in hBN as a source of single defect emission.

\newpage

\section{Evolution of defect states from ab-initio computations}

%
%
\begin{figure}[!ht]
\scalebox{\figurescale}{\includegraphics[width=1.7\linewidth]{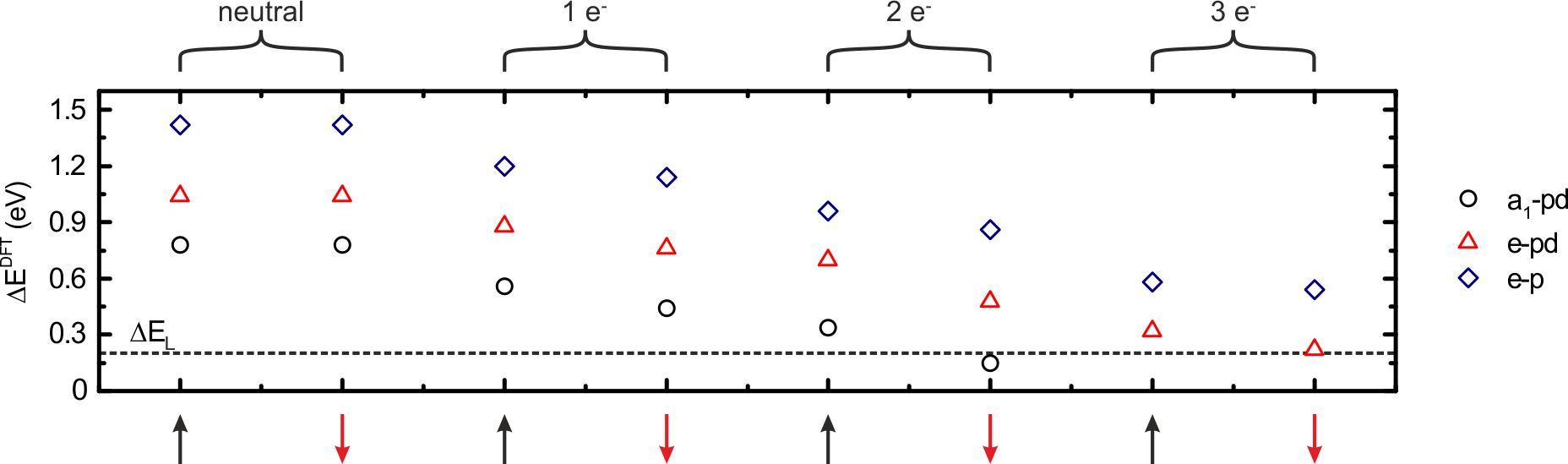}}
\renewcommand{\figurename}{SI Fig.}
\caption{\label{SIfigstates}
\textbf{Energy difference of ingap defect states for differently charged molybdenum-vacancies.}
Ab-initio computed energy differences $\Delta E^{DFT} = E(CBM)^{DFT} - E(state)^{DFT}$ are shown for a neutral $V^{0}_{Mo}$ and for a single, double and triple charged molybdenum-vacancy ($V^{1-}_{Mo}$, $V^{2-}_{Mo}$ and $V^{3-}_{Mo}$). Energy differences are shown for the $e$-$p$, $e$-$pd$ and $a_{1}$-$pd$ for both, spin down and spin up states. The experimentally obtained lowest emission energy of $\Delta E_{L} \sim \SI{0.2}{\electronvolt}$ is highlighted with a dashed line.
}
\end{figure}

SI Fig.~\ref{SIfigstates} presents the DFT calculated energy differences $\Delta E^{DFT}$ as taken from the DOS shown in Fig.~1f-i in the main manuscript. In particular, energy differences between the conduction band minimum (CBM) and the $e$-$p$, $e$-$pd$ and $a_{1}$-$pd$ spin up and spin down states are shown for a neutral $V^{0}_{Mo}$ and for a single, double and triple charged molybdenum-vacancy ($V^{1-}_{Mo}$, $V^{2-}_{Mo}$ and $V^{3-}_{Mo}$). Moreover, we highlight the experimentally observed highest energy detuning $\Delta E_{L} \sim \SI{0.2}{\electronvolt}$ in order to directly compare with theory. We find best agreement with the $e$-$pd$ spin down state ($\Delta E^{DFT} \sim \SI{0.22}{\electronvolt}$) of the $V^{3-}_{Mo}$ that also is the lowest unoccupied state for this charging configuration.

\newpage

\section{Quantum emitter lineshape}

To describe the quantum emitter spectrum we utilize the independent boson model that has been succcessfully applied to describe the lineshape of quantum dot states~\cite{zimmermann_dephasing_2002,wilson-rae_quantum_2002,krummheuer_theory_2002} and defect-bound excitons~\cite{duke_phonon-broadened_1965}. The Hamiltonian for localized excitons of energy $E$ coupled to lattice vibrations reads
\begin{align}
    H=EX^\dagger X + \sum_{\mathbf q} \hbar\omega_{\mathbf q} a^\dagger_{\mathbf q}a_{\mathbf q} + X^\dagger X \sum_{\mathbf q} g^X_{\mathbf q}(a^\dagger_{\mathbf q} + a_{-{\mathbf q}})\,,
\end{align}
where the interaction part represents lattice deformations due to the presence of an exciton. The exciton-phonon matrix elements 
\begin{align}
    g^{X}_{\mathbf q} = \sum_{\mathbf k} \phi^*({\mathbf k})\left[g^c_{\mathbf q}\phi({\mathbf k}+{\mathbf q}_h)-g^v_{\mathbf q}\phi({\mathbf k}-{\mathbf q}_e)\right]
\end{align}
contain the exciton wave function $\phi({\mathbf k})$ with ${\mathbf q}_{\text{e/h}}$ been the electron/hole momentum. For a localized $s$-exciton we obtain
\begin{align}
    g^{X}_{\mathbf q}=\frac{g^c_{\mathbf q}}{\left[1+\frac{(a_{\text{B}}|{\mathbf q}_h|)^{2}}{4}\right]^2}-\frac{g^v_{\mathbf q}}{\left[1+\frac{(a_{\text{B}}|{\mathbf q}_e|)^{2}}{4}\right]^2}
\end{align}
where $a_{\text{B}}$ is the 2D exciton bohr radius. The carrier-phonon matrix elements $g^{c/v}$ are treated in deformation potential approximation~\cite{kaasbjerg_phonon-limited_2012, jin_intrinsic_2014} and we account for the coupling with LA and TA phonon modes, which are found to be the dominant source of phonon dephasing in the system. For deformation potential coupling with acoustic phonons of dispersion $\omega_{\mathbf q} = v_s q$ the carrier-phonon matrix element is given by
\begin{align}
    g^{c/v}_{\mathbf q} = D^{c/v} \sqrt{\frac{\hbar\omega_{\mathbf q}}{2v_s^2\rho A}}
\end{align}
with sound velocity $v_s$, 2D mass density $\rho$, and deformation potentials $D^{c/v}$ taken from DFT/DFPT calculations~\cite{li_intrinsic_2013, jin_intrinsic_2014}. Here, $A$ is the crystal area.

The spectral properties are obtained from the optical response to a classical electric field $E(t)$, which drives a polarization of the medium. In response to a weak optical (test) field, the linear optical susceptibility is given by:
\begin{align}
    \chi(\omega) = \frac{P(\omega)}{\varepsilon_0 E(\omega)}\,.
    \label{eq:chi}
\end{align}
considering the polarization $P$ in direction of the external field. To obtain the frequency-dependent response of the medium, we use the Fourier transform of the polarization $P(t)$, which can be written for a delta-like excitation as~\cite{zimmermann_dephasing_2002}
\begin{align}
    P(t) = -i\Theta(t)\exp{\left[ -it(E - i\gamma_{\text{rad}}) + R(t) - R(0) - tR'(0) \right]}\,.
    \label{eq:P_t}
\end{align}
The polarization decays due to radiative recombination with a rate $\gamma_{\text{rad}}$ as well as excitations into phonon sidebands described by the complex function
\begin{align}
    R(t) = \sum_{\mathbf q} \left|g^X_{\mathbf q}\right|^2 \left[\frac{N_{\mathbf q}}{(\omega_{\mathbf q} + i\gamma_{\text{ph}})^2} e^{(+i\omega_{\mathbf q} - \gamma_{\text{ph}})t}+\frac{N_{\mathbf q} + 1}{(\omega_{\mathbf q} - i\gamma_{\text{ph}})^2} e^{(-i\omega_{\mathbf q} - \gamma_{\text{ph}})t}\right]
    \label{eq:R_t}
\end{align}
where $N_{\mathbf q} = 1/(\exp[\hbar\omega_{\mathbf q}/k_BT]-1)$ is the phonon occupation, $g^X_{\mathbf q}$ the exciton-phonon matrix elements, $\gamma_{ph}$ the phonon lifetime and $\omega_{q}$ the acoustic phonon dispersion assumed in equilibrium at lattice temperature $T$. Absorption spectra that are obtained from solving Eqs.~\eqref{eq:chi}-\eqref{eq:R_t} posses a Lorentz-broadend zero-phonon line as well as phonon sidebands due to multi-phonon processes. The latter give rise to the observed asymmetric lineshape at low temperature, where only phonon emission processes dominate. 

\begin{figure}[t]
    \includegraphics[width=0.5\textwidth]{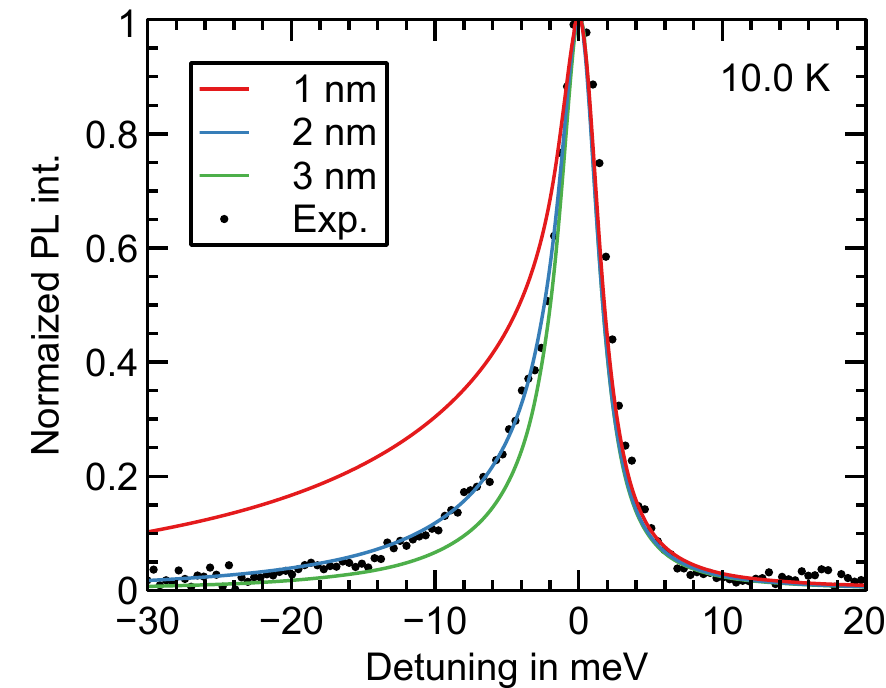}
    \renewcommand{\figurename}{SI Fig.}
    \caption{\textbf{Comparison between measured and calculated emission spectra for various Bohr radii.} The result for $a_{\text{B}}=2\,$nm corresponds to the spectrum at $\SI{10}{\kelvin}$ in Fig.~3 of the main text.}
    \label{fig:theo_exp_cmp}
\end{figure}
Emission spectra are obtained as a mirror image of the absorption spectra reflected across the zero-phonon line~\cite{mahan_many_2000}. At low temperature best agreement between calculated and measured photoluminescence spectra is obtained for a Bohr radius of $\SI{2}{\nano\meter}$, see SI Fig.~\ref{fig:theo_exp_cmp}. In order to obtain this agreement, it was necessary to reduce the exciton-phonon coupling strength by a factor of $3.75$ compared to the bare monolayer. This results in a Huang-Rhys factor 
\begin{align}
    S = \sum_{\mathbf q} \frac{\left|g^X_{\mathbf q}\right|^2}{\omega_{\mathbf q}^2}
\end{align}
of $0.75$, which is consistent with the phonon coupling that we have obtained from the analysis of the temperature dependent peak position (cf. main text). For the phonon lifetime $1/\gamma_{\text{ph}}$ we used a value of $\SI{40}{\pico\second}$ according to Ref.~\cite{gu_layer_2016}. From the fit we obtain a value for the radiative linewidth of $\SI{0.5}{\milli\electronvolt}$. Using this parameters, all spectra in Fig.~3 of the main text are then well described by changing the temperature according to the experimental condition.

%
%
%
%
\bibliographystyle{achemso}
\bibliography{full}